\documentclass[onecolumn,showpacs]{revtex4}

\usepackage{psfig}
\usepackage{graphicx}
\usepackage{dcolumn}
\usepackage{amsmath}

\makeatletter
\def\btt#1{\texttt{\@backslashchar#1}}%
\DeclareRobustCommand\bblash{\btt{\@backslashchar}}%
\makeatother

\begin{document}

\title{Global Monopole in Asymptotically dS/AdS Spacetime}

\author{Xin-zhou Li}\email{kychz@shtu.edu.cn}
\author{Jian-gang Hao}

\affiliation{ Shanghai United Center for Astrophysics, Shanghai
Normal University, Shanghai 200234 ,China
}%

\date{\today}

\begin{abstract}
Abstract : In this paper, we investigate the global monopole in
asymptotically dS/Ads spacetime and find that the mass of the
monopole in the asymptotically dS spacetime could be positive if
the cosmological constant is greater than a critical value. This
shows that the gravitational field of the global monopole could be
attractive or repulsive depending on the value of the cosmological
constant.
\end{abstract}

\pacs{11.27.+d, 98.80.Cq}

\maketitle

\vspace{1cm}

Various of kinds of topological defects could be produced by the
phase transition in the early Universe and their existence has
important implications in cosmology\cite{vilenkin}. Global
monopole, which has divergent mass in flat spacetime, is one of
the most important above mentioned defects. The property of the
global monopole in curved spacetime, or equivalently, its
gravitational effects, was firstly studied by Barriola and
Vilenkin\cite{barriola}. When one considers the gravity, the
linearly divergent mass of global monopole has an effect analogous
to that of a deficit solid angle plus that of a tiny mass at the
origin. Harari and Loust\`{o}\cite{harari}, and Shi and
Li\cite{li1} have shown that this small gravitational potential is
actually repulsive. A new class of cold stars, addressed as
D-stars(defect stars) have been proposed by Li et.al.\cite{li2,
li3,li4}. One of the most important features of such stars,
comparing to Q-stars, is that the theory has monopole solutions
when the matter field is abscent, which makes the D-stars behave
very differently from the Q-stars. The topological defects are
also investigated in the Friedmann-Robertson-Walker
spacetime\cite{Basu}.On the other hand, there has been a renewed
interest in AdS spacetime due to the theoretical speculation of
AdS/CFT correspondence, which state that string theory in anti-de
Sitter space (usually with extra internal dimensions) is
equivalent to the conformal field theory in one less
dimension\cite{maldacina, witten}. Recently, the holographic
duality between quantum gravity on de Sitter(dS) spacetime and a
quantum field theory living on the past boundary of dS spacetime
was proposed\cite{strominger} and the vortices in dS spacetime was
studied by Ghezelbash and Mann\cite{mann}. Many authors
conjectured that the dS/CFT correspondence bear a lot of
similarities with the Ads/CFT correspondence, although some
interpretive issues remain. The monopole and dyon solution in
gauge theories based on the various gauge group have been
found\cite{hooft}. However, in flat space there can not be static
soliton solution in the pure Yang-Mills theory\cite{deser}. The
presence of gravity can supply attractive force which binds
non-Abelian gauge field into a soliton. The cosmological constant
influence the behavior of the soliton solution significantly. In
asymptotically Minkowski spacetime the electric components are
forbidden in static solution\cite{volkov}. If the spacetime
includes the cosmological constant, forbidding the electric
components of the non-Abelian gauge fields fail, thus allowing
dyon solutions. A continuum of new dyon solutions in the
Einstein-Yang-Mills theory in asymptotically AdS spacetime have
been investigated\cite{bjoraker}, which are regular everywhere and
specified with their mass, and non-Abelian electric and magnetic
charges. Similarly, the presence of cosmological constant affects
the behavior of the global monopole remarkably. If the spacetime
is modified to include the positive cosmological constant, the
gravitation field of global monopole can be attractive in contrast
to the same problem in asymptotically Minkowski or AdS spacetime.

In this paper, we study the global monopole in asymptotically
AdS/dS spacetime and show that the mass of the monopole might be
positive in asymptotically dS spacetime if the cosmological
constant is greater than a critical value.

\vspace{0.4cm}
 \vspace{0.4cm}

The Lagrangian for the global monopole is
\begin{equation}\label{lag}
L=\frac{1}{2}\partial_\mu\phi^a\partial^\nu\phi^a-\frac{1}{4}\lambda^2(\phi^a\phi^a-\sigma_0)^2
\end{equation}

\noindent where $\phi^a$ is the triplet of Goldstone field and
possesses a internal O(3) symmetry. When the symmetry breaks down
to U(1), there will exist topological defects known as monopole.
The configuration describing monopole solution is

\begin{equation}\label{config}
  \phi^a=\sigma_0f(\rho)\frac{x^a}{\rho}
\end{equation}

\noindent where $x^ax^a=\rho^2$ and $a=1,2,3.$

When $f$ approaches unity at infinity, we will have a monopole
solution. The static spherically symmetric metric is

\begin{equation}\label{metric}
  ds^2=B(\rho)dt^2-A(\rho)d \rho^2-\rho^2(d\theta^2+ \sin^2\theta d\varphi^2)
\end{equation}

By introducing the dimensionless parameter $r=\sigma_0 \rho$, we
obtain the equations of motion for goldstone field as:

\begin{equation}\label{scalareq}
  \frac{1}{A}f^{''}+[\frac{2}{Ar}+\frac{1}{2B}(\frac{B}{A})^{'}]f^{'}-\frac{2}{r^2}f
  -\lambda^2(f^2-1)f=0
\end{equation}

\noindent where the prime denotes the derivative with respect to
$r$.

In dS/AdS spacetime, the Einstein equation is
\begin{equation}\label{einsteineq}
G_{\mu\nu}+\beta g_{\mu\nu}=\kappa T_{\mu\nu}
\end{equation}

\noindent where $\beta$ is the cosmological constant and $\kappa=8
\pi G$. dS and AdS spacetime correspond to the cases that $\beta$
is positive and negative respectively. The Einstein equations in
dS/AdS spacetime now are ready to written as:

\begin{equation}\label{einsteineq1}
 -\frac{1}{A}(\frac{1}{r^2}-\frac{1}{r}\frac{A^{'}}{A})+\frac{1}{r^2}=\epsilon^2T_0^0
 -\frac{\beta}{\sigma_0^2}
\end{equation}

\begin{equation}\label{einsteineq2}
 -\frac{1}{A}(\frac{1}{r^2}+\frac{1}{r}\frac{B^{'}}{B})+\frac{1}{r^2}=\epsilon^2T_1^1
 -\frac{\beta}{\sigma_0^2}
\end{equation}
\noindent where

\begin{eqnarray}\label{energytensor}
 T_0^0=D+U+V\\
 T_1^1=D+U-V
\end{eqnarray}

\noindent are energy momentum tensors and
\begin{eqnarray}\label{defination}
 D&&=\frac{f^2}{r^2}\nonumber\\
 U&&=\frac{\lambda^2}{4}(f^2-1)^2\\
 V&&=\frac{f^{'2}}{2A}\nonumber
\end{eqnarray}

\noindent and $\epsilon^2=\kappa\sigma_0^2$ is a dimensionless
parameter. Solving Eqs.(\ref{einsteineq1}) and
(\ref{einsteineq2}), one can obtain

\begin{equation}\label{solution1}
 A^{-1}(r)=1-\epsilon^2+\frac{\beta}{3\sigma_0^2}r^2-\frac{2G\sigma_0M_A(r)}{r}
\end{equation}

\begin{equation}\label{solution2}
 B(r)=1-\epsilon^2+\frac{\beta}{3\sigma_0^2}r^2-\frac{2G\sigma_0M_B(r)}{r}
\end{equation}

\noindent where

\begin{eqnarray}\label{ma}
M_A(r)=4\pi\sigma_0\exp[-\triangle(r)]\times\int_0^r
dy\exp[\triangle(y)]
\{f^2-1+y^2[U+(1-\epsilon^2+\frac{\beta}{3\sigma_0^2}y^2)f^{'2}]\}
\end{eqnarray}

\noindent and

\begin{eqnarray}\label{mb}
M_B(r)=M_A(r)\exp[\widetilde{\triangle}(r)]
+\frac{r(1-\epsilon^2+\frac{\beta}{3\sigma_0^2}r^2)}{2}\{1-\exp[\widetilde{\triangle}(r)]\}
\end{eqnarray}

\noindent In which
\begin{equation}\label{delta}
\triangle(r)=\frac{\epsilon^2}{2}\int_0^rdy(yf^{'2})
\end{equation}

\noindent and

\begin{equation}\label{delta}
\widetilde{\triangle}(r)=\epsilon^2\int_\infty^rdy(yf^{'2})
\end{equation}

Next, we discuss the behavior of these functions in asymptotically
dS/AdS spacetime. A global monopole solution $f$ should approaches
unity when $r\gg1$. If this convergence is fast enough then
$M_A(r)$ and $M_B(r)$ will also quickly converge to finite values.
Therefore, we can find the asymptotic expansions:
\begin{equation}\label{asymf}
  f(r)=1-\frac{3\sigma_0^2}{\beta+3\lambda^2\sigma_0^2}\frac{1}{r^2}-
  \frac{9[2\beta\epsilon^2\sigma_0^4+3(2\epsilon^2-3)\lambda^2\sigma_0^6]}
  {2(2\beta-3\lambda^2\sigma_0^2)(\beta+3\lambda^2\sigma_0^2)^2}\frac{1}{r^4}+O(\frac{1}{r^6})
\end{equation}

\begin{equation}\label{asyma}
  M_A(r)=M_A(\beta, \epsilon^2)+\frac{4\pi\sigma_0}{r}+O(\frac{1}{r^3})
\end{equation}

\begin{equation}\label{asymb}
  M_B(r)=M_A(r)(1-\frac{\epsilon^2}{r^4})+\frac{4\pi\sigma_0(1-\epsilon^2)}{r^3}+O(\frac{1}{r^7})
\end{equation}

\noindent where $M_A(\beta, \epsilon^2)\equiv \lim _{r\rightarrow
\infty}M_A(r)$, which is a function dependent on $\beta$ and
$\epsilon^2$. The dependence on $\epsilon$ of asymptotic behavior
is quite weak for the global monopole solution. However, the
asymptotic behavior is evidently dependent of the parameters
$\beta$, $\sigma_0$ and $\lambda$. The
Eqs.(\ref{asymf})-(\ref{asymb}) in the limit of small cosmological
constant reduce to the well known ones in flat spacetime. The
solution Eqs.(\ref{solution1})-(\ref{solution2}) induce a deficit
angle in asymptotically dS/AdS spacetime. Numerical calculation
for $f(r)$ show that its shape is quite insensitive to $\epsilon$
in the range $0\leq\epsilon\leq 1$ not only asymptotically, but
also close to the origin. We also find that an increasing positive
cosmological constant tends to make a thicker monopole solution
and a decreasing negative cosmological constant tends to make a
thinner monopole solution.

In the following, We present a numerical analysis to the system
and the results are shown in Fig.1.

\psfig{figure=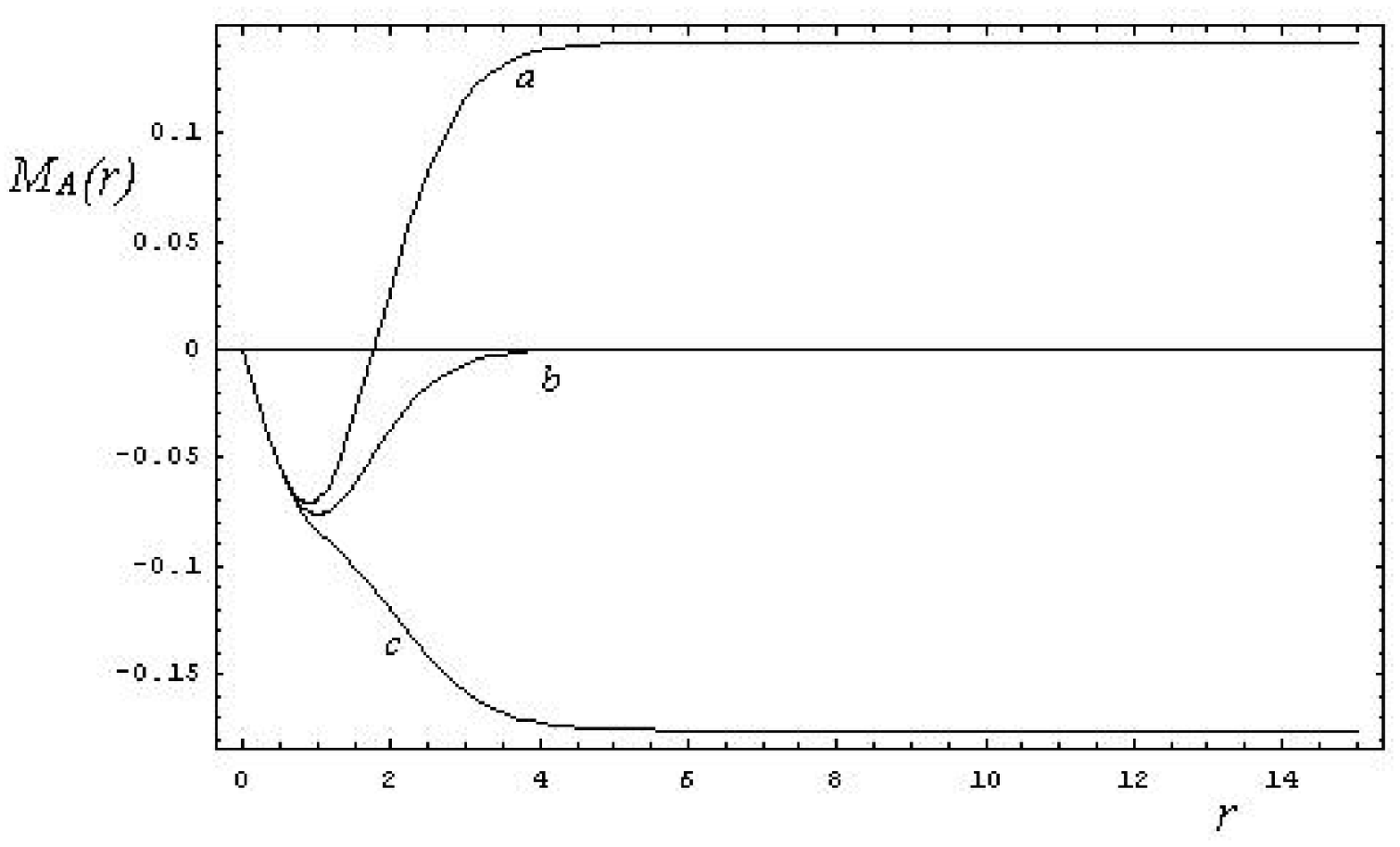,height=3.5in,width=5in}
 Fig1. The plot of mass as a function of $r$. Here we choose
 $\lambda=1$, $\sigma_0=0.01$, $G=1$. Curve (a), (b) and (c)
 are plotted when $\beta=0.0010, 0.0003$ and $-0.0005$ respectively.

\hspace{3cm}

From Fig.1, one can find that the mass decrease to a negative
asymptotic value when $r$ approaches infinity in AdS spacetime.
But in dS spacetime, the mass will be positive if the cosmological
constant is large enough.  The critical value for the cosmological
constant is 0.0003 in our set-up. The asymptotic mass for the
above Curve (a), (b) and (c) are 0.1415, 0.0000 and -0.1760
respectively. It is clear that the presence of cosmological
constant affects the behavior of the global monopole
significantly. If the spacetime is modified to include a positive
cosmological constant, the gravitational field of global monopole
can be attractive in contrast to the same problem in
asymptotically flat or AdS spacetime.

Finally, We want to discuss why the tiny mass of global monopole
is defined as the limiting value of the function $M_{A}(r)$ given
by Eq.(\ref{ma}). The standard definition of the ADM mass is
different: It is defined by the
$g_{rr}^{-1}=1-\frac{2GM(r)}{r}-\frac{\lambda}{3}r^2$, where
$\lambda$ is the cosmological constant. The mass, $M$, is then the
limiting value of $M(r)$, and it is always positive. This agrees
with the well-known positive mass theorem for the dS/AdS space
firstly proved by \cite{abbott}. However, in the case of global
monopole, $M(r)$ is linearly divergent for large $r$, which leads
to a deficit solid angle plus a residual effect of gravitational
field. When a test particle moves in the gravitational field of
the global monopole, the repulsive or attractive nature of the
residual gravitational effect can be perceived by this
particle\cite{harari,li1}. That is, the standard definition of the
mass gives in the case of a global monopole a linearly divergent
expression plus a constant term. Now, it turns out that the
divergent term does not produce any gravitational effect on the
matter interacting with the monopole, and the whole interaction is
entirely determined by the subleading finite term. This is why it
is customary to subtract the divergent term from the definition of
mass, since it is the resulting finite `effective' mass that
determines the gravitational interaction with the monopole.
Furthermore, the attractive or repulsive property of the residual
gravitational field is determined by the positiveness or
negativeness of $M_A(r)$. For the relation between the Positive
Mass Conjecture and global monopole, one may refer to
Ref.\cite{cvetic1}, in which Cveti\^{c} and Soleng pointed out
that "In Ref\cite{cvetic2}, a Positive Mass Conjecture was
formulated saying that there is no singularity free solution of
Einstein's field equations with matter sources(not including the
vacuum) obeying the weak energy condition equations for which an
exterior observer can see a negative mass object. A global
monopole Ref\cite{barriola,harari,li1} would appear to be a
counter example, but in this case the Goldstone fields extended to
infinity, which means that these objects are extended sources and
all observers must be inside the system". Therefore, although the
effective mass $M_A$ is negative under certain circumstance, it
will not contradict the Positive Mass Conjecture\cite{cvetic2} and
the Positive Mass theorem in dS/AdS spacetime\cite{abbott}.

When one study the motion of test particles around a global
monopole, it is an excellent approximation to take $M_A(r)$ as the
constant $M_A(\beta, \epsilon^2)$ since the effective mass
approaches its asymptotic value very quickly. Apart from the
academic interest in the global monopole configuration, the
D-stars\cite{li2,li3,li4} seem to make it relevant to astronomical
situation. Work on the generalization of D-stars to asymptotically
ds/Ads spacetime is in preparation.

\vspace{0.8cm} \noindent ACKNOWLEDGMENTS

This work was partially supported by National Nature Science
Foundation of China under Grant No. 19875016, National Doctor
Foundation of China under Grant No. 1999025110, and Foundation of
Shanghai Development for Science and Technology under Grant No.
01JC1435.

\end{document}